\documentclass[aps,showpacs,twocolumn]{revtex4}
\usepackage{amsfonts}
\usepackage{amsmath}
\usepackage{amssymb}
\usepackage{graphicx}
\usepackage{bm}

\input{tcilatex}

\begin{document}

\title{Polarization of active Janus particles}
\author{Thomas Bickel$^{1}$, Guillermo Zecua$^{2}$, and Alois W\"{u}rger$%
^{1} $}
\affiliation{$^{1}$Laboratoire Ondes et Mati\`{e}re d'Aquitaine, Universit\'{e} de
Bordeaux \& CNRS, 33405 Talence, France}
\affiliation{$^{2}$Institut f\"{u}r Theoretische Physik, Universit\"{a}t Leipzig, 04103
Leipzig, Germany }

\begin{abstract}
We study the collective motion of Janus particles in a temperature or
concentration gradient $\mathbf{\nabla }\psi $. Because of the torque
exerted by an external or self-generated field, the particles align their
axis on the gradient $\mathbf{\nabla }\psi $. In a swarm of self-driven
particles, this polarization enhances the interaction-driven confinement.
Self-polarization in a non-uniform laser beam could be used for guiding hot
particles along a given trajectory.

\end{abstract}

\maketitle

Swarms of self-propelling birds, fish, or insects show dynamical patterns
that arise from the fact that each individual adapts to the state of its
neighbors \cite{Vic12}. Similar phenomena occur for liquid dispersions of
active objects.\ Thus cells of \textit{Escherichia coli} move along a
chemical gradient generated by their neighbors and form stable spatial
structures \cite{Adl66}.\ More recently, self-driven Janus particles (JPs)
have been studied as a model active system \cite{Pax05,Gol05}. It has been
shown that the interplay of self-propulsion and rotational diffusion leads
to Brownian motion with an enhanced diffusion coefficient \cite%
{How07,Jia10,Pal10,Vol11,But12,Bar13}, and that the particles' chemical
activity results in cluster formation \cite{The12,Pal13,But13} and oriented
motion \cite{Bar13a}. Guiding a single microswimmer along a given trajectory
can be achieved by dynamical feedback \cite{Qia13}.

Active colloids have been realized by partly coating silica or polystyrene
particles with a metal or carbon layer; self-propulsion arises from
non-uniform surface properties such as temperature or chemical activity. An
excess temperature $\psi =T$ is induced by heating the metal or carbon cap
through absorption of laser light \cite{Jia10,Vol11,But12,But13,Qia13} or
magnetic fields \cite{Bar13}. Chemical signalling with a molecular solute, $%
\psi =c$, is achieved by electrocatalysis of hydrogen peroxide at a metal
cap \cite{Pax05,How07,The12,Pal13}.

A minimal model for self-driven systems consists in a drift-diffusion
equation that was originally designed for motile bacteria with chemotactic
interactions \cite{Kel70}, and that describes complex spatial structures
observed in cell cultures \cite{Bre98}. More recently this model was adapted
to JPs that aggregate due to the chemical gradient generated by their
electrocatalytic activity \cite{The12}. For thermally active colloids,
similar results were derived from the Smoluchowski equation \cite%
{Gol12,Coh13}: Self-propulsion strongly enhances the diffusion term, whereas
the drift velocity arises from the gradient field $\nabla \psi $ generated
by the neighbor particles; a sufficiently strong attractive drift term may
even cause the implosion of a swarm of JPs.

In the present Letter we show that active colloids are polarized by their
chemical or thermal interactions \cite{preprint}. A non-uniform field $\psi $
exerts a viscous torque on the JP, which in turn aligns its symmetry axis
along the gradient $\nabla \psi $. Self-propulsion of such oriented JPs adds
a novel contribution to the drift velocity, which is dominant for strong
driving or large P\'{e}clet number and which may lead to novel collective
effects. It turns out that this oriented self-propulsion corresponds to the
usual model for bacteria motility.

\begin{figure}[ptb]
\includegraphics[width=\columnwidth]{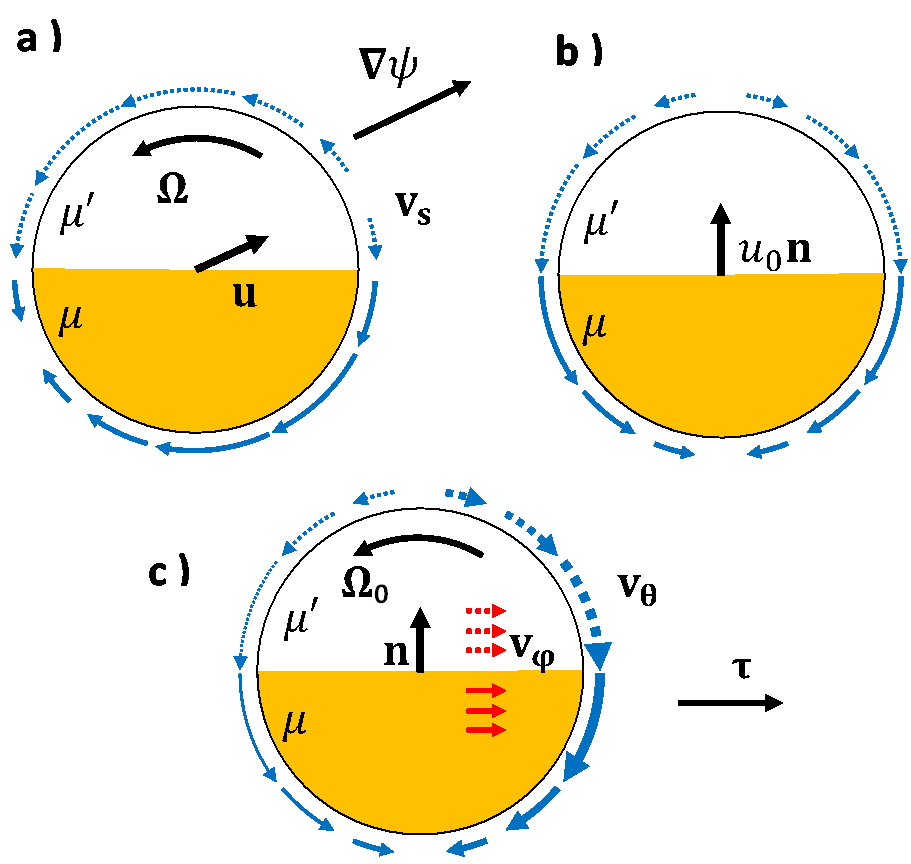}
\caption{a) Motion of a Janus particle resulting from an external field gradient 
$\mathbf{\protect\nabla }\protect\psi $. The particle moves at velocity $%
u\propto \protect\mu ^{\prime }+\protect\mu $ along the field gradient and
rotates at rate $\Omega \propto \protect\mu ^{\prime }-\protect\mu $. The
mobility takes the values $\protect\mu ^{\prime }$ on the insulating part
and $\protect\mu $ on the metal cap; the quasi-slip velocities are indicated
by full and dotted arrows.\ b) Self-propulsion of an active JP. The
particle's own field $\protect\psi _{S}$ leads to a quasislip velocity $%
\mathbf{v}_{s}=\protect\mu \protect\nabla _{\Vert }\protect\psi _{S}$ that
is symmetric with respect to the particle's axis. The particle self-propels
at a velocity $u_{0}\propto \protect\mu ^{\prime }+\protect\mu $. c)
Non-uniform source field with gradient $\mathbf{\protect\tau }$. The
quasislip velocity at the particle surface has contributions in polar and
azimuthal directions; both the magnitude of $\mathbf{v}_{\protect\theta }$
and the orientation of $\mathbf{v}_{\protect\varphi }$ are sensitive to the
intensity gradient $\mathbf{\protect\tau }$. The resulting angular velocity $%
\mathbf{\Omega }_{0}$ contains terms proportional to $\protect\mu ^{\prime
}\pm \protect\mu $, as given in (\protect\ref{18}).}
\end{figure}

\textit{Polarization of Janus particles}. Consider a JP interacting with a
concentration or temperature field $\psi $. Within a thin boundary layer,
the parallel component of the gradient of the local field $\bar{\psi}$
induces a quasislip velocity along the particle surface \cite{And89,Gol07} 
\begin{equation}
\mathbf{v}_{s}(\mathbf{r})=\mu (\mathbf{r})\mathbf{\nabla }_{\parallel }\bar{%
\psi}(\mathbf{r}).  \label{1}
\end{equation}%
The main result of this paper arise from the material-dependent non-uniform
mobility constant $\mu $ and from the properties of the local field $\bar{%
\psi}$. We consider the case where two values $\mu $ and\ $\mu ^{\prime }$
occur on the two halfspheres of an otherwise homogeneous JP, as indicated in
Fig.\ 1. The resulting quasislip velocities are indicated as solid and
dotted lines; their sign and magnitude depend on the mobility values and on
the local gradient $\mathbf{\nabla }_{\parallel }\bar{\psi}$.

The quasi-slip velocity (\ref{1}) constitutes the boundary condition for the
velocity field $\mathbf{v(r)}$ of the surrounding fluid, $\mathbf{v|}_{B}%
\mathbf{=u+\Omega \times r}_{B}+\mathbf{v}_{s}$, where $B$ indicates the
outer limit of the interaction layer, typically at a few nanometers from the
particle surface.\ This relation determines the linear and angular
velocities of the JP. The former is given by the surface average $\mathbf{u}%
=-S^{-1}\int dS\mathbf{v}_{s}$, and reads explicitly \cite{suppmat} 
\begin{equation}
\mathbf{u}=-\xi _{1}\frac{\mu +\mu ^{\prime }}{3}\mathbf{\nabla }\psi 
\mathbf{(r)}.  \label{8}
\end{equation}%
Depending on the sign of $\mu +\mu ^{\prime }$, the particles move along or
opposite to the field gradient. For uniform surface properties ($\mu =\mu
^{\prime }$) one recovers the usual phoretic velocity of particles in an
external field $\mathbf{\nabla }\psi $ \cite{And89}.

The factor $\xi _{1}$ in\ (\ref{8}) accounts for the deformation of the
field $\psi $ due to the presence of the JP. The local gradient is given by
the projection on the tangential plane, $\mathbf{\nabla }_{\parallel }\bar{%
\psi}=\xi _{1}(1-\mathbf{\hat{r}\hat{r})\cdot \nabla }\psi $, with the
surface normal $\mathbf{\hat{r}}$. For constant $\mathbf{\nabla }\psi $ the
correction factor reads $\xi _{1}=3\kappa _{s}/(2\kappa _{s}+\kappa _{p})$;
in the case of temperature $\kappa _{s}$ and $\kappa _{p}$\ are the heat
conductivities of solvant and particle \cite{suppmat}. This form is valid
for a sufficiently thin cap, such that the metal layer does not modify the
heat flow pattern. In contrast, a thick metal layer results in a spatially
varying $\xi (\mathbf{r})$, reducing the slip velocity on the cap and
enhancing it on the insulating hemisphere \cite{Bic13}; the limiting case of
an isothermal cap is accounted for by putting $\mu =0$ and augmenting $\mu
^{\prime }$ by a factor that depends on the orientation of the\ JP. If $\psi 
$ is an electric potential, $\kappa _{s}$ and $\kappa _{p}$ are the
permittivities; for dielectric materials in water one has $\xi _{1}\approx 
\frac{3}{2}$ In the case of an applied concentration gradient the $\kappa
_{i}$ are the diffusion coefficients; if the solute does not penetrate the
particle, one has $\xi =1$; additional effects occur for chemical reactions 
\cite{Sah13}.

A non-uniform mobility factor $\mu (\mathbf{r})$ gives rise to a rotational
component of the quasislip velocity \cite{Sah13, suppmat}. For example, for $%
\mu ^{\prime }=0$ the dotted arrows in Fig.\ 1a vanish, and the remaining $%
\mathbf{v}_{s}$ results in a clockwise motion of the surrounding fluid; in
order to minimize the viscous stress, the particle then turns in the
opposite direction until its axis is parallel to $\mathbf{\nabla }\psi $.
Equilibrating the surface and viscous forces one obtains the angular
frequency 
\begin{equation}
\mathbf{\Omega }=-\frac{3}{2a}\oint \frac{d\mathbf{S}\times \mathbf{v}_{s}}{%
4\pi a^{2}}\mathbf{,}  \label{10}
\end{equation}%
where $d\mathbf{S}$ is the oriented surface element and $a$\ the particle
radius. Performing the integral and introducing the unit vector $\mathbf{n}$
along the particle axis, one finds 
\begin{equation}
\mathbf{\Omega }=\mathbf{n}\times \mathbf{A,\ \ \ \ \ A}=-\frac{3\xi
_{1}(\mu ^{\prime }-\mu )}{8a}\mathbf{\nabla }\psi \mathbf{(r)}.  \label{14}
\end{equation}%
Note that the angular velocity is proportional to the mobility difference $%
\mu -\mu ^{\prime }$\ of the two hemispheres and thus vanishes for a
homogeneous surface. Yet in general $\mu $ and $\mu ^{\prime }$ are quite
different from each other.\ If the heated metal cap forms an isotherm, its
thermophoretic mobility $\mu $ is zero \cite{Bic13}, whereas $\mu ^{\prime }$
may take either sign depending on the precise driving mechanism \cite%
{Vig10,Wue10}. On the other hand, an electrocatalytic mechanism is described
finite $\mu $ at the Pt cap and $\mu ^{\prime }=0$ at the remaining surface.

The viscous stress underlying (\ref{10}), tends to orient the JP along the
external field, whereas rotational diffusion with coefficient $D_{r}$ favors
dispersion. The resulting kinetics are described by the Smoluchowski
equation for the distribution function $f(\mathbf{n})$, 
\begin{equation}
\partial _{t}f=-\mathcal{R}\cdot \left( \mathbf{\Omega }-D_{r}\mathcal{R}%
\right) f\equiv \mathcal{L}_{\mathbf{n}}f,  \label{15}
\end{equation}%
with the rotation operator $\mathcal{R}=\mathbf{n\times \nabla }_{\mathbf{n}%
} $, and the gradient $\mathbf{\nabla }_{\mathbf{n}}$ with respect to the
orientation of the JP \cite{Gol12}. This equation is readily solved in terms
of the angle $\theta $ between the particle axis and the field gradient,
resulting in $\Omega =A\sin \theta $. The corresponding equation for the
steady-state, $(A\sin \theta +D_{r}\partial _{\theta })f=0$, is readily
solved, $f(\theta )\propto e^{(A/D_{r})\cos \theta }$. This effective
rotational potential aligns the JP axis on the field gradient, with the mean
orientation 
\begin{equation}
\mathbf{n}_{\text{eq}}=\left( \coth \frac{A}{D_{r}}-\frac{D_{r}}{A}\right) 
\frac{\mathbf{A}}{A}.  \label{30}
\end{equation}%
For micron-size particles in a temperature gradients of the order K/$\mu $m,
the ratio $A/D_{r}$ may exceed unity, which is confirmed by experiments on
thermophoretic traps with $ua/D>1$ \cite{Leo09}. In other words, in such
systems JPs are perfectly aligned, $n_{\text{eq}}=1$, whereas for weak
fields one has $n_{\text{eq}}=A/3D_{r}$. Polarization in an external field $%
\psi _{\text{ext}}$ is illustrated in Fig. 2a).

\begin{figure}[ptb]
\includegraphics[width=\columnwidth]{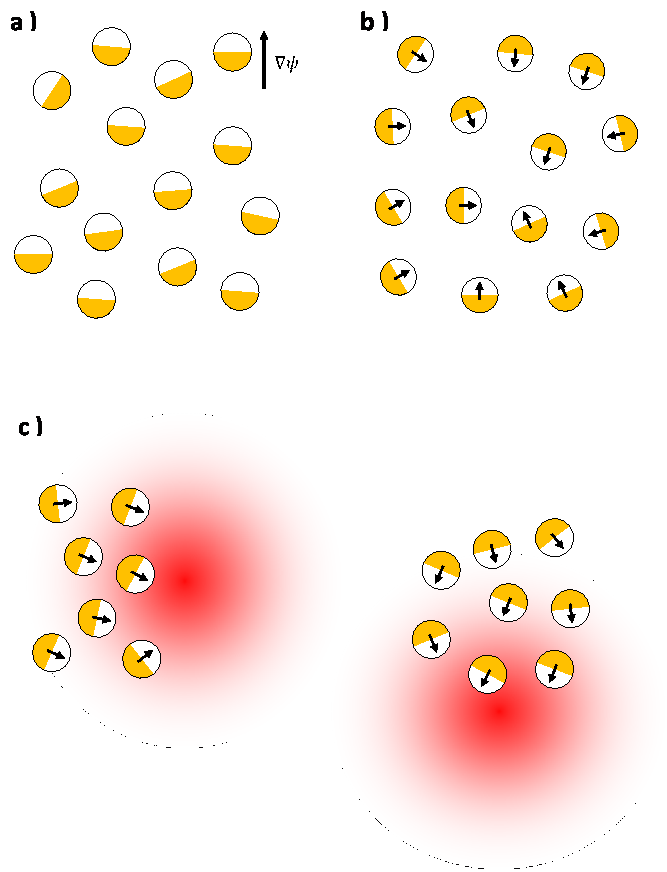}
\caption{Polarization of Janus particles in a field gradient $%
\mathbf{\protect\nabla }\protect\psi $. a) Case of an externally applied
field $\protect\psi $; the orientation of the particles is given by Eq. (%
\protect\ref{30}).\ There is no self-propulsion, $u_{0}=0$; the small
translational velocity $\mathbf{u}$ is not indicated.\textbf{\ }b)\textbf{\ }%
Self-confinement of a swarm of Janus particles. With an appropriate choice
of the mobilities $\protect\mu $ and $\protect\mu ^{\prime }$, the drift
velocity (\protect\ref{28}) points toward the center of the swarm, thus
favoring cluster formation or even implosion \protect\cite{Gol12}. The
polarization is due to the interaction field $\protect\psi $ which in the
simplest case has radial symmetry and its related to the density by $\protect%
\nabla ^{2}\protect\psi +k\protect\rho =0$. c) Guided self-propulsion.
Self-driven hot JPs are polarized either by the beam intensity gradient $%
\mathbf{\protect\tau }$ of the heating laser, or by an external temperature
gradient $\protect\nabla T_{\text{ext}}$ that results from local heating of
the solvant with an IR\ laser. The JPs follow the heated spot, first to the
right and then downward in the figure. }
\end{figure}

\textit{Self-propelling Janus particles}.\ Now we consider a swarm of active
JPs as shown in Fig.\ 2b). Their motion consists of single-particle and
interaction contributions: Each particle self-propels in its own non-uniform
field $\psi _{S}$, whereas that of the neighbors, $\psi \mathbf{(\mathbf{r})=%
}\sum_{j}\psi _{j}\mathbf{(\mathbf{r-r}}_{j}\mathbf{)}$, results in the
linear and angular velocities (\ref{8}) and (\ref{14}).

The self-generated term $\psi _{S}$ arises from the active surface property $%
Q$; for example, the surface temperature $T_{S}$ is modified by laser
heating at power $Q=\beta I(\mathbf{r}_{0})$, where $I$ is the beam
intensity at the particle position $\mathbf{r}_{0}$, and $\beta $ the
absorption coefficient per unit area of the cap.\ For constant power $Q_{0}$%
, the quasislip velocity is symmetric about the particle axis, as
illustrated in Fig.\ 1b, and results in self-propulsion at a speed $\mathbf{n%
}u_{0}=S^{-1}\oint dS\mu (\mathbf{r})\mathbf{\nabla }_{\parallel }\psi _{S}$%
.\ Solving the diffusion equation for $\psi _{S}$ one finds \cite{suppmat} \ 
\begin{equation}
u_{0}=\xi _{1}\frac{\mu +\mu ^{\prime }}{3}\frac{Q_{0}}{8\kappa _{s}},
\end{equation}%
where $Q_{0}/\kappa _{s}$ gives the mean gradient of\ $\psi _{S}$.

For a spatially varying laser intensity $I(\mathbf{r})$ (or concentration of
a catalytic agent \cite{Sah13}), the source field breaks the axial symmetry.
The resulting slip velocity has both polar and azimuthal components, as
illustrated in Fig.\ 1c), and exerts a viscous torque 
\begin{subequations}
\label{18}
\begin{equation}
\mathbf{\Omega }_{0}=\mathbf{n\times A}_{0}
\end{equation}%
which is perpendicular on the particle axis $\mathbf{n}$ and on the gradient
of the source, $\mathbf{\tau }=\mathbf{\nabla }\ln I(\mathbf{r}_{0})$, 
\begin{equation}
\mathbf{A}_{0}=-\mathbf{\tau }\left( \left( \mu ^{\prime }-\mu \right) \frac{%
5}{18}\xi _{1}+\left( \mu ^{\prime }+\mu \right) \widetilde{\xi }\right) 
\frac{3Q_{0}}{8\kappa _{s}}.
\end{equation}%
The first term involves the mobility difference and the dipolar deformation
factor $\xi _{1}$.\ The second one, proportional to $\mu ^{\prime }+\mu $,
arises from even-order multipoles of $\psi _{S}$ and carries a correction
factor $\widetilde{\xi }$ which is given by the weighted average of the
diffusivity contrast factors $\xi _{n}=(2n+1)/(n+1+n\kappa _{p}/\kappa _{s})$
\cite{suppmat}. For $\mu ^{\prime }>\mu $ both terms result in
counterclockwise rotation as shown in Fig.\ 1c.) The case of catalytic
activity is discussed in \cite{Sah13}.

Comparing self-propulsed motion with driving due to neighbors in a swarm of
JPs, we find that for a colloidal volume fraction of at most a few percent,
the velocity $u_{0}$ is much larger than that due to an external field, $u$.
The torques exerted by the field of an active neighbor at distance $R$ and
by a intensity gradient vary as $\Omega \sim u_{0}a/R^{2}$ and $\Omega
_{0}\sim u_{0}\tau $, respectively; depending on the system parameters, one
or the other may dominate.

The state of a given JP is described by its position $\mathbf{r}$ and the
orientation of its axis $\mathbf{n}$. The distribution function $P(\mathbf{%
r,n})$ obeys the equation 
\end{subequations}
\begin{equation}
\partial _{t}P=-\mathbf{\nabla }\cdot \left( u_{0}\mathbf{n}+\mathbf{u}-D%
\mathbf{\nabla }\right) P+\mathcal{L}_{\mathbf{n}}P.  \label{2}
\end{equation}%
The first term on the right-hand side describes translational motion with
velocity $u_{0}\mathbf{n}+\mathbf{u}$\ and gradient diffusion with Einstein
coefficient $D$. The second term accounts for rotational motion according to
(\ref{15}); the diffusion coefficients are related through $D=\frac{4}{3}%
a^{2}D_{r}$.

An approximate solution of (\ref{2}) is obtained by inserting the moment
expansion $P(\mathbf{r,n})=\rho (\mathbf{r})+\mathbf{n\cdot p}(\mathbf{r}%
)+...$, integrating over $\mathbf{n}$, and truncating the resulting
hierarchy at finite order. Following Golestanian \cite{Gol12}, we neglect
quadrupolar contributions and other small terms in the equation for the
polarization vector $\mathbf{p(r)}=(1/4\pi )\int d\mathbf{nn}P$, and thus
find \cite{suppmat} 
\begin{equation}
\mathbf{p}=-\frac{u_{0}}{6D_{r}}\mathbf{\nabla }\rho +\mathbf{n}_{\text{eq}%
}\rho \mathbf{,\ \ \ n}_{\text{eq}}=\frac{\mathbf{A+A}_{0}}{3D_{r}}.
\label{24}
\end{equation}%
The first term, which has been derived in previous work \cite{Gol12,Cat10},
accounts for the diffusive transport of polarization in a non-uniform
density; the prefactor $u_{0}/D_{r}$ gives the distance over which the
particle self-propels during its rotational relaxation time; with $u_{0}\sim
10$ $\mu $m/sec and $1/D_{r}\sim 1$ sec one finds about 10 microns. The
second term $\mathbf{n}_{\text{eq}}\rho $ accounts for active polarization
of JPs. Fig. 2b) illustrates the alignement on the field gradient (\ref{14})
created by nearby JPs, and Fig.\ 2c) shows self-polarization of the JPs
along the gradient (\ref{18}).

With the polarization $\mathbf{p}$ one obtains the drift-diffusion equation
for the density $\rho $ \cite{suppmat} 
\begin{equation}
\partial _{t}\rho =-\mathbf{\nabla }\cdot \left( \mathbf{u}_{\text{eff}}\rho
-D_{\text{eff}}\mathbf{\nabla }\rho \right) ,  \label{26}
\end{equation}%
where $D_{\text{eff}}=D(1+\frac{2}{9}\mathrm{Pe}^{2})$ is the effective
diffusion coefficient and $\mathrm{Pe}=u_{0}a/D$ the P\'{e}clet number \cite%
{Gol05,How07}. The effective velocity 
\begin{equation}
\mathbf{u}_{\text{eff}}=\mathbf{u+}u_{0}\mathbf{n}_{\text{eq}}  \label{27}
\end{equation}%
consists of the interaction-driven drift (\ref{8}) and oriented
self-propulsion with the equilibrium polarization $\mathbf{n}_{\text{eq}}=(%
\mathbf{A+A}_{0})/3D_{r}$.

In a constant source field $Q$ there is no self-polarization, $\mathbf{A}%
_{0}=0$, and the drift velocity can be cast in the form 
\begin{equation}
\mathbf{u}_{\text{eff}}=\xi \left( -\frac{\mu +\mu ^{\prime }}{3}+\mathrm{Pe}%
\frac{\mu -\mu ^{\prime }}{6}\right) \mathbf{\nabla }\psi \mathbf{(r).}
\label{28}
\end{equation}%
The first term in parentheses, which has derived previously \cite%
{The12,Gol12}, is independent of the particle orientation. The second one
has not been considered so far; it arises from self-propulsion of polarized
JP and dominates at large P\'{e}clet number. Since $\mathrm{Pe}\propto \mu
+\mu ^{\prime }$, the two terms in (\ref{28}) carry opposite signs for $\mu
>\mu ^{\prime }$.

Temperature and concentration fields generated by the JPs' heat absorption
or chemical activity, satisfy $\mathbf{\nabla }^{2}\psi +k\rho =0$ with the
particle density as source term.\ Then the sign of the prefactor of $\mathbf{%
u}_{\text{eff}}$ determines whether self-propulsion disperses or confines a
cloud of JPs. A sufficiently large negative drift velocity results in
clustering as illustrated in Fig. 2b \cite{The12,Pal13,But13} and may even
drive implosion of the swarm \cite{Gol12}. This latter scenario has been
discussed in detail for $\mathbf{u}_{\text{eff}}=\mathbf{u}$ and $\mu =\mu
^{\prime }<0$, corresponding to a negative Soret coefficient \cite{Gol12}.
The correction term $u_{0}\mathbf{n}_{\text{eq}}$ derived here, is dominant
for \TEXTsymbol{\vert}$\mathrm{Pe|}>0$ and, according to (\ref{28}) results
in attraction $\mu ^{\prime 2}>\mu ^{2}$, independently of the sign of the
mobilities. Thus polarizaton enhances $\mathbf{u}_{\text{eff}}$ by a factor $%
\mathrm{Pe}$ and, at large P\'{e}clet number, even modifies the dynamical
phase diagram. Experiments on cluster formation \cite{The12,Pal13,But13} and
oriented motion \cite{Bar13a} support the qualitative features of the
drift-diffusion model, yet available data are not sufficiently precise for a
quantitative comparison.

\textit{Guided self-propulsion.}\ So far we discussed polarization along the
field gradient $\nabla \psi $ generated by the heat absorption or chemical
activity of neighbor JPs. Here we discuss the case where both propulsion and
polarization result from the particle's self-generated temperature field $%
T_{S}$.\ With the linear velocity $u_{0}$ and the order parameter $\mathbf{n}%
_{\text{eq}}=\mathbf{A}_{0}/3D_{r}$, we obtain oriented motion along the
intensity gradient of the laser beam, 
\begin{equation}
\mathbf{u}_{\text{eff}}=u_{0}\mathbf{n}_{\text{eq}}=\frac{4}{9}\mathrm{Pe}a%
\mathbf{A}_{0}.  \label{32}
\end{equation}%
Note that this a single-particle property and varies with the square of the
laser intensity. A physical realization is sketched in Fig.\ 2c, where a
focussed laser beam illuminates a swarm of JPs. Since the particles move
towards the center of the beam according to (\ref{10}),\ they could be
guided by a mobile laser beam along a given trajectory.

\textit{Chemotaxis of bacteria.} We compare the motion of polarized JPs with
bacteria that are guided by chemotactic signalling. \textit{E. coli}
self-propels through flagella rotating in the \textquotedblleft
run\textquotedblright\ mode at a velocity $u_{0}$ along its axis $\mathbf{n}$
\cite{Bre98}. After a period of $\tau \sim 1$ sec, they switch to the
\textquotedblleft tumble\textquotedblright\ mode, which randomly changes the
orientation and thus plays the role of rotational diffusion. The cell
performs a random walk with diffusion coefficient $D_{\text{eff}}\sim
u_{0}^{2}\tau $.

Bacteria are not able to actively reorient in a field gradient, contrary to
JPs according to (\ref{10}). Yet they are sensitive to the concentration of
certain solutes. If a cell detects a favorable change of $\psi $ along its
trajectory, it augments the time $\tau $; on the other hand, if it feels it
goes the wrong direction, it switches more rapidly to the tumble mode.\ As a
consequence, the bacterium spends more time in an orientation toward the
source \cite{Bre98}. Assuming a linear variation with the concentration
gradient, one has $\tau =\tau _{0}+\alpha \mathbf{n\cdot \nabla }\psi $,
where $\alpha $ describes the strength of the response to chemical
signalling. The resulting polarization $\mathbf{n}_{\text{eq}}=\frac{1}{3}%
\alpha \mathbf{\nabla }\psi $ results in the drift velocity 
\begin{equation}
\mathbf{u}_{\text{eff}}=u_{0}\mathbf{n}_{\text{eq}}=\frac{u_{0}}{3}\alpha 
\mathbf{\nabla }\psi .  \label{34}
\end{equation}%
Comparison with the drift velocity of JPs shows that bacteria motion
corresponds to the second term in Eq. (\ref{27}), that is, to
self-propulsion along the fieldd gradient $\mathbf{\nabla }\psi $.\ 

In view of Eqs.(\ref{27}) and (\ref{34})\ one expects for swarms of JPs a
dynamical behavior very similar to that observed in bacteria cultures.
Fine-tuning of the surface parameters $\mu $ and $\mu ^{\prime }$\ would
allow to separate the effects of phoretic motion $\mathbf{u}$\ and of
oriented self-propulsion $u_{0}\mathbf{n}_{\text{eq}}$. Since only the
latter is present in (\ref{34}), the relative weight of these terms is an
important parameter when comparing the motion of JPs and bacteria.

We conclude with a remark on hydrodynamic interactions which have been
neglected in the present paper. The interactions considered here are
mediated by thermal or concentration gradients $\mathbf{\nabla }\psi $ which
in three dimensions vary with the square of the inverse distance, $\mathbf{%
\nabla }\psi \propto r^{-2}$. Depending on the symmetry of quasi-slip
velocity, hydrodynamic interactions decay as $r^{-3}$ or $r^{-2}$ \cite%
{Bic13,Llo10,Zoe13}; the latter term may attain values comparable to the
interaction contribution $\mathbf{u}$ in (\ref{27}). Yet at large P\'{e}clet
number, it is small as compared to the self-propulsion contribution $u_{0}%
\mathbf{n}_{\text{eq}}$.

\textbf{Acknowledgment.} A.W. acknowledges support through the Leibniz
program of Universit\"{a}t Leipzig during the summer term 2013, and thanks
the groups of Frank Cichos and Klaus Kroy for their kind hospitality.

\pagebreak

\end{document}